\documentclass[runningheads]{llncs}
\usepackage{graphicx}
\usepackage{multirow}
\usepackage{adjustbox}
\usepackage{lipsum}
\usepackage{algorithmic}
\usepackage{graphicx}
\usepackage{float}
\usepackage{url}
\usepackage{colortbl}
\usepackage{threeparttable}
\usepackage{tcolorbox}
\usepackage{arydshln}

\begin{document}

\title{Kuksa\textsuperscript{*}: Self-Adaptive Microservices in Automotive Systems}

\titlerunning{Kuksa\textsuperscript{*}: Self-Adaptive Microservices in Automotive Systems}

\author{Ahmad Banijamali\inst{1}\orcidID{0000-0002-6283-142X} \and
Pasi Kuvaja\inst{1}\orcidID{[0000-0002-1488-6928} \and
Markku Oivo\inst{1}\orcidID{0000-0002-1698-2323} \and Pooyan Jamshidi\inst{2}\orcidID{0000-0002-9342-0703}}
\authorrunning{Banijamali et al.}

\institute{M3S Research Unit, ITEE Faculty, University of Oulu, Finland \email{\{firstname.lastname\}@oulu.fi}\\ \and
Computer Science and Engineering Department, University of South Carolina, USA\\
\email{pjamshid@cse.sc.edu}}
\maketitle

\begin{abstract}
In pervasive dynamic environments, vehicles connect to other objects to send operational data and receive updates so that vehicular applications can provide services to users on demand.\ Automotive systems should be self-adaptive, thereby they can make real-time decisions based on changing operating conditions.\ Emerging modern solutions, such as microservices could improve self-adaptation capabilities and ensure higher levels of quality performance in many domains.\ We employed a real-world automotive platform called Eclipse Kuksa to propose a framework based on microservices architecture to enhance the self-adaptation capabilities of automotive systems for runtime data analysis.\ To evaluate the designed solution, we conducted an experiment in an automotive laboratory setting where our solution was implemented as a microservice-based adaptation engine and integrated with other Eclipse Kuksa components.\ The results of our study indicate the importance of design trade-offs for quality requirements' satisfaction levels of each microservices and the whole system for the optimal performance of an adaptive system at runtime.
\vspace{-.15cm}

\end{abstract}

\vspace{-.6cm}

\section{Introduction}
\vspace{-.15cm}

The importance of software in the automotive domain has increased because most of the innovations in new cars originate from software, large computing power, and modern sensors.\ A vehicle is now a part of a network to collaborate with other vehicles, cloud, edge, and the surrounding infrastructure to deliver value-added services \cite{nobre2019vehicular}.\ Cloud and edge platforms have enabled new services based on access to car data and functions from outside the car.\ Because of the large volume of these data, service management in the domain of connected vehicles is appealing for reliable and efficient solutions \cite{shukla2018software}, allowing for scalable and autonomic processing.

In the highly dynamic environment that automotive systems operate, services should be able to automatically and quickly adapt to changes at runtime to preserve a level of quality under the complex operating conditions, such as scaling workloads and faults \cite{mendoncca2019developing}.\ To turn vehicles into self-adaptive systems, we have to combine self-managing capabilities with architectural design patterns that enable self-adaptive functioning under complex working conditions \cite{banijamalisoftware}.

Microservice architectures are a potential solution for designing and deploying self-adaptive business capabilities \cite{newman2015building}.\ Although there have been many attempts in the development of self-adaptive systems, a small number of microservices solutions, such as Kubernetes, load balancers, and circuit breakers could find their ways to the industry to facilitate autonomic management of microservice systems \cite{mendoncca2019developing}.

There is also an increasing interest in microservices in the automotive domain; however, the research on the microservice architectures that enable self-adaptive automotive systems is still in the early stage.\ A previous survey \cite{banijamali2020software} highlighted this research gap for architecture-based self-adaptation in automotive systems.\ To address this need, we empirically investigated how microservices patterns, such as load balancing, resiliency mechanisms, and monitoring systems can improve the self-adaptive quality performance under the dynamic conditions of the automotive domain \cite{banijamali2019kuksa}.\ In this paper, we extend the study on the microservice architectures in the automotive domain \textit{to suggest a framework called Kuksa\textsuperscript{*} that aims at improving the self-adaptation capabilities of automotive systems by proposing a microservice-based autonomic controller.}\ We employed the Eclipse Kuksa architecture\footnote{https://www.eclipse.org/kuksa/} in the automotive domain to design Kuksa\textsuperscript{*} and evaluated it in an automotive experimental laboratory setting.

The results of this study are valuable for the system architects and academic researchers in the automotive domain.\ Industrial practitioners can benefit from experimental results on the performance and configurations of self-adaptive microservice systems in a real-world industrial case in the automotive domain.\ Academics can also gain insights into quality research gaps, service optimisation, and quality trade-off challenges in self-adaptive systems.\ The key contributions of the study are as follows: (1) the Kuksa\textsuperscript{*} framework to enhance the self-adaptation capabilities of automotive systems by using microservice architectures and (2) the empirical evaluation of the framework in an experimental setting and showing the results of design and runtime trade-offs.

\section{Background and Related Work}
\label{sec2}

This section provides a brief overview of previous studies and related work.
\vspace{-.2cm}

\subsection{Background}
\label{sec2A}
\vspace{-.1cm}

\subsubsection{Microservices:}
\label{sec2A1}
Large monolithic systems hardly can scale when different modules have conflicting resource requirements, life cycles, and deployment frameworks \cite{banijamali2019kuksa}.\ Thus, there has been an increasing interest to migrate systems to more scalable and reliable architectures, such as microservices \cite{bass2015devops}.\ Breaking down applications into a set of smaller, interconnected services trades external complexity for microservice simplicity \cite{newman2015building}.

Microservices are a design alternative that is used to foster further research on self-adaptive systems \cite{mendoncca2019developing}.\ They isolate business functions into small services to optimise the autonomy, modifiability, and replaceability of the services.\ Microservice architectures pertain to challenges that should be carefully addressed.\ For example, they increase the complexity due to only the fact that they are distributed.\ Therefore, it is necessary to have inter-process and autonomic communications among different microservices \cite{bass2015devops}.\ Although microservices improve the communication with the back-end databases, they apply partitioned database architecture, which means it is often necessary to update multiple databases that are owned by different services \cite{pahl18} and to have an eventual consistency-based approach.\ Also, a large number of microservices in an application may increase latency and performance issues \cite{bass2015devops}.\ Therefore, microservices should work as self-adaptive systems that dynamically learn and adapt their behaviours to preserve specific quality levels under dynamic operating conditions \cite{mendoncca2019developing}.
\vspace{-.18cm}

\subsubsection{Cloud platforms in the automotive domain:}
\label{sec2A2}
Cloud platforms in the automotive domain are designed for high scalability and flexibility and offer the continuous delivery of new software applications, such as telematics, infotainment, fleet management, and remote diagnostics and maintenance \cite{shiftmobility}.\ The creation of scalable, extensible, and hardware-agnostic platforms has facilitated the deployment of collaborative vehicle applications \cite{Staron}.\ With the vehicle movement, it is necessary not only to have access to an infinitely scalable architecture supporting data analytics and application development in the cloud \cite{siegel2017survey} but also to adapt to changes in the operating conditions \cite{sun2018implementation}.\ For instance, when there is an interruption in the communication network, services in vehicles or the cloud should adopt alternative offline scenarios \cite{sun2018implementation}.
\vspace{-.15cm}

\subsection{Related work}
\label{sec2B}
\vspace{-.1cm}

Architecture-based adaptation is addressed by (1) the adaptation capability of a system to change with minimal human intervention and (2) the control loop mechanisms in the system that separate core services from adaptation services \cite{weyns2018applying}.\ Policy-based architecture adaptation is an approach to map a functional situation to relevant action, strategy, or reconﬁguration \cite{ho2015model}.\ We used this approach to decouple a system adaptation logic with knowledge about how to react when an adaptation is required.\ Architecture-based adaptation allows architectural reconfiguration based on predeﬁned rules \cite{weyns2018applying}.

Architectural configurations impact the performance of self-adaptive software systems \cite{pahl17}.\ There exist many techniques, such as Bayesian optimisation \cite{jamshidi2016uncertainty} and multi-objective optimisation \cite{filieri2015automated} that have been used for moving towards the optimal performance of self-adaptive systems.\ Furthermore, many approaches, such as planning and constraint-solving algorithms have been used to find a new target configuration that satisfies the given constraints of a system \cite{zeller2012timing}.

Control loops are commonly used in policy-based architecture adaptation approaches to collect data about external environments and make autonomic decisions \cite{Kephart}.\ In this regard, inducing microservices with the primitives of self-adaptivity is a strong candidate for addressing autonomic trade-offs \cite{Hasan}.\ Pereira et al.\ \cite{Pereira} presented an adaptive system that designed the control loop components in the form of microservices that are easily deployed in a container-based system, such as Kubernetes.\ Another study \cite{Aderaldo} proposed an architecture-based self-adaptation service for cloud-native applications based on the Rainbow self-adaptation framework with support for Docker containers and Kubernetes.\ Current approaches mainly rely on the quantitative verification of system properties, including techniques to produce formal guarantees about the quantitative aspects of systems, such as performance \cite{eberhardinger2018measuring}.

There has been considerable work in the area of managing software systems by using service compositions and adaptations \cite{delemos}.\ However, they hardly address the problem from the dynamic adaptation perspective by balancing microservices performance at runtime, specifically in the automotive domain.\ Our study has tackled this issue and investigated how self-adaptive microservices can optimise automotive systems' performance at runtime.

\vspace{-.1cm}

\section{Research Method}
\label{sec3}
\vspace{-.1cm}

This section describes the current study's objective, research questions, research methods, and the Eclipse Kuksa framework.
\vspace{-.2cm}

\subsection{Objective and research questions}
\label{sec3A}
\vspace{-.1cm}

The study objective is to design a microservice architecture for enhancing the self-adaptation capabilities of automotive systems.\ To achieve this objective, we devised the following research questions (RQs):

\begin{tcolorbox}
\begin{itemize}
\item RQ1: How can microservices be applied in the design of self-adaptive systems in the automotive domain?
\item RQ2: What are the quality trade-offs at runtime when using self-adaptive microservices in the automotive domain? 
\end{itemize}
\end{tcolorbox}
\vspace{-.25cm}

\subsection{Research methods}
\label{sec3B}
\vspace{-.1cm}

To propose the Kuksa\textsuperscript{*} framework, we applied the concept of control loops (MAPE-K) \cite{Kephart} to render a design solution that improves over the system life cycle through accumulating knowledge.\ In addition, we opted for an experimental method and a laboratory setting (see section \ref{sec5} for the detailed description of the experimental scenario and setting) that allowed us to investigate performance trade-offs from the Kuksa\textsuperscript{*} framework and the system under evaluation regarding self-adaptation capabilities.\ We investigated how self-adaptation microservices can adapt the system to changes and how they can register changes into the whole system.\ The laboratory setting helped us to demonstrate and control the impact of changes on the performance of the designed architecture because real context evaluations are often safety-critical, complex, and costly~\cite{a2000experimentation}.

We designed multiple microservices to enable the self-adaptation in our designed setting and to evaluate the system performance based on an \textit{``adaptive video streaming scenario''}.\ We selected this scenario because of its importance in other automotive scenarios, such as the driver assistance, collision-avoidance, safety-critical scenarios, vehicle-to-vehicle communication, autonomous driving, and comfort systems and its significance in security scenarios in the automotive domain.\ The video streaming presents an example scenario to show the performance of self-adaptive microservices under highly dynamic conditions of automotive systems, although Kuksa\textsuperscript{*} has high flexibility to be adopted and customised for other scenarios in the automotive domain.

We used the results of previous studies (\cite{eberhardinger2018measuring,McGeoch}) to create a metric for measuring the performance of the self-adaptive system under evaluation and comparing it with the static system architecture.\ The performance of software systems denotes its capabilities in its execution \cite{eberhardinger2018measuring}, in which two important measures include the solution quality and the time taken to achieve the solution \cite{McGeoch}.\ Hence, the total performance in our designed system composed of time performance and solution quality performance \cite{eberhardinger2018measuring}, as follows: \[p(sys) = wt.tp(sys) + wq.qp(sys)\] where $wt+wq = 1$.\ The parameters of $wt$ and $wq$, respectively, indicated the weight factors (importance) for time performance $tp(sys)$ and quality performance $qp(sys)$.\ Normalisation of $tp(sys)$ and $qp(sys)$ in a range of 0 to 1 allowed us to sum them up in our metric.

Time and quality performances were measured based on the average performance from multiple runs.\ The $tp(sys)$ used the average time required for the reconfiguration of the video streaming microservice which translates into the time involved in finding a new adaptation strategy configuration in a separate run $r$.\ The $tp(sys)$ was calculated based on the ratio of the time needed for reconfiguration in a run $r$, as follows: \[tp(r) = 1-\frac{\sum Reconfiguration~time(r)}{Duration(r)}\] The quality performance was calculated based on the achieved video stream quality, using two factors of \textit{frame rate} and \textit{frame quality}, that the system was able to apply within a run $r$, as follows: \[qp(r) = \frac{\sum Quality(r)}{Quality~max}\] The codomain of qp(r) was [0, 1] and a value close to 1 indicated a better-achieved quality as 1 would imply that the maximum quality performance was~reached.
\vspace{-.1cm}

\subsection{Eclipse Kuksa}
\label{sec3C}
\vspace{-.1cm}

\begin{figure*}[t]
\centering
\scalebox{0.36}{\includegraphics{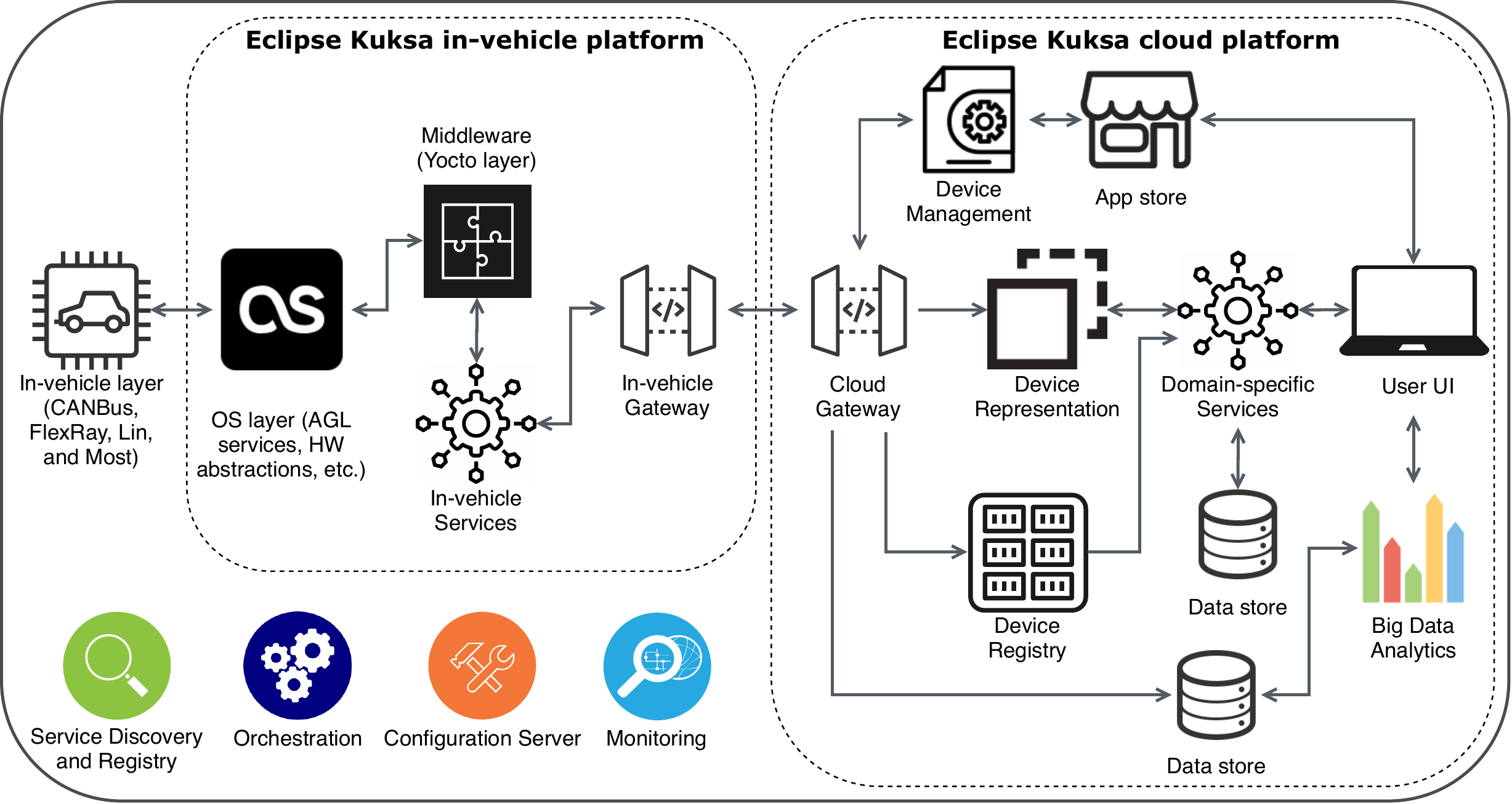}}
\vspace{-.27cm}
\caption{Software architecture of Eclipse Kuksa.}
\vspace{-.7cm}
\label{fig1}
\end{figure*}

Eclipse Kuksa is a project that provides an open and secure cloud platform to connect a wide range of vehicles to the cloud via in-car and Internet connections.\ It aims at the mass differentiation of vehicles through application systems, software solutions, and services.\ The project comprises three open-source software platforms for the (1) in-vehicle, (2) cloud, and (3) application integrated development environment, shown in Figure~\ref{fig1} (adopted from \cite{banijamali2019kuksa}).

\textit{Gateways (in-vehicle/cloud)} enable data communication from vehicles (i.e., car CAN-Bus) or control commands delivery to vehicles.\ The gateways provide remote service interfaces for connecting vehicles and devices to the cloud back-end using protocols, such as MQTT.\ The \textit{middleware layer} in the in-vehicle platform includes APIs to abstract the vehicle's electrical/electronic architecture and communication libraries to enable communication services and manage the network access.\ \textit{OS layer} includes, for example, AGL core services, boot loader, hardware abstractions, and platform update manager.

\textit{Device representation} provides a digital representation of devices to realise the distinct functionality of domain-specific services.
\textit{Device registry} grants access to a distinct functionality for only eligible devices and users.
\textit{Domain-specific services} are developed according to various use cases and scenarios in the cloud or vehicles to handle different functions.
\textit{Data storage} is necessary for reliable data management systems that can handle the data complexities related to the size, consistency, performance, scalability, and security of different microservices.
\textit{Device management} is responsible for tasks, such as the provisioning, configuration, monitoring, and diagnostics of connected devices.
\textit{Kuksa app store} is a digital repository for software applications developed by vehicle manufacturers and other third-party providers.
\textit{Visualisation and data analytics} efficiently identify, collect, clean, analyse, and visualise that data to enable new services.

The Kuksa architecture \cite{banijamali2019kuksa} relies on more microservices, including \textit{Service discovery and registry}, which automatically registers and deregisters service instances.\ It provides a source to find out which of the service instances are currently available.
\textit{Configuration server} stores service configurations to isolate the configuration properties from codes and to enable autonomous service rebuilding or restarting.
\textit{Monitoring} presents the state of microservices-based systems using consolidated logs, reports, and infrastructure-level metrics, for example, for monitoring service usage and finding performance bottlenecks.
\textit{Orchestration} collectively provides the mechanisms that deploy, maintain, and scale services and applications based on configuration settings to meet different workloads.

\section{Kuksa*: A Self-Adaptive Microservice-based Framework for Automotive Systems}
\label{sec4}

Automotive systems, such as driver assistance or safety-critical systems should allow for modifications of services and systems by taking appropriate actions at runtime based on changes in the vehicle operating condition \cite{delemos}.\ The actions can include, for example, replacing one microservice instance with another (e.g., upgrading power train and safety electronic services based on the new trailer attached to the vehicle), changing the number of microservice replicas (e.g., increasing load of vehicular data communication with surrounding objects in a crowded area), and dynamically changing the quality requirements of microservices (e.g., adding new safety requirements and constraints in bad weather conditions) \cite{mendoncca2019developing}.\ The functions are automated at runtime in a control loop that collects the details from automotive systems and sensors and acts accordingly.

\begin{figure}[t]
\centering
\scalebox{0.43}{\includegraphics{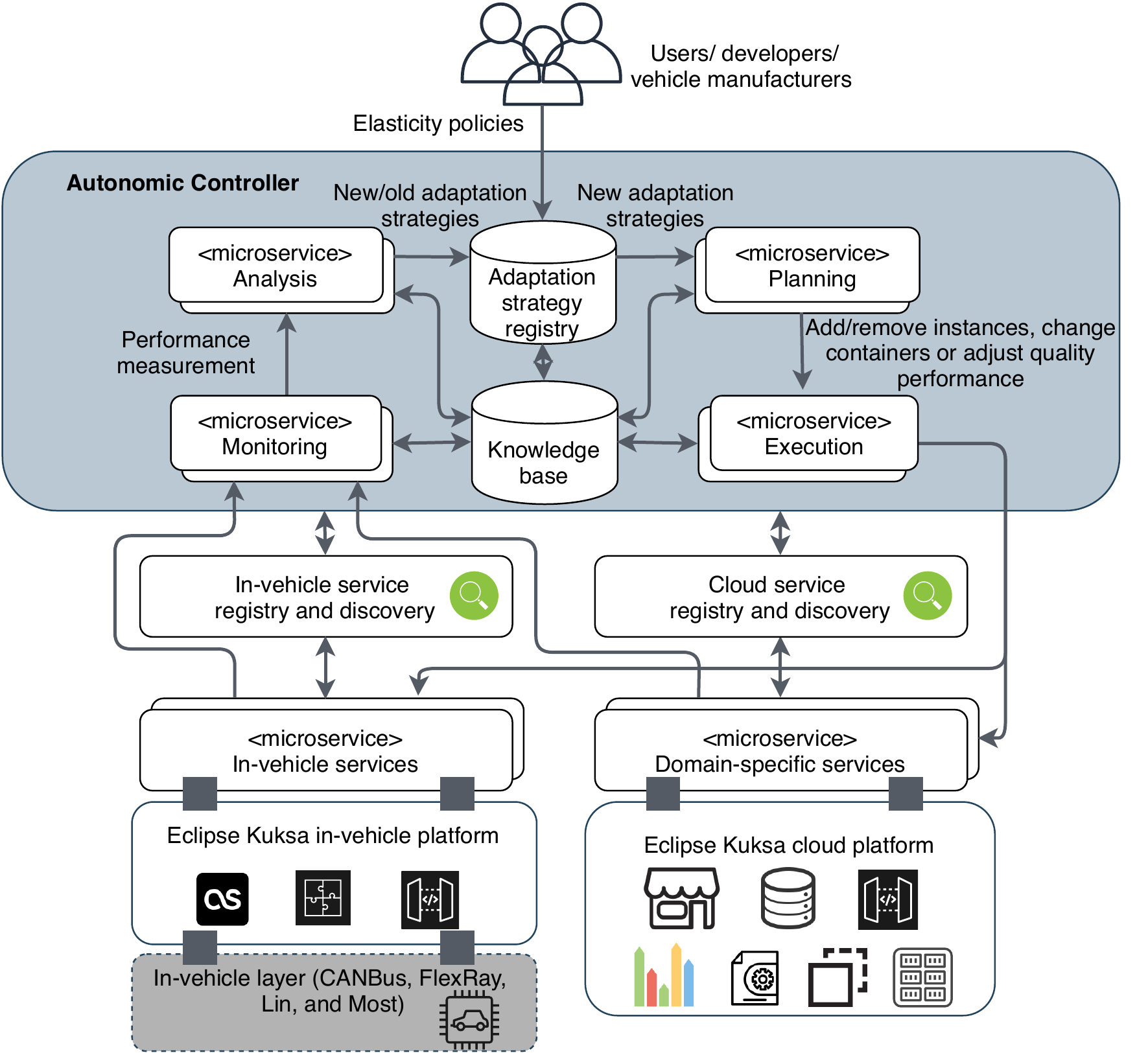}}
\vspace{-.35cm}
\caption{The Kuksa\textsuperscript{*} framework.}
\vspace{-.8cm}
\label{fig2}
\end{figure}

The autonomic controller, shown in Figure~\ref{fig2}, applies the microservice architecture to improve the in-vehicle and cloud systems with self-adaptation capabilities.\ It continuously monitors and analyses vehicle's working conditions and executes appropriate actions to resolve issues or improve system quality performance in vehicles with minimal human intervention.\ Kuksa\textsuperscript{*} was constructed on top of the Eclipse Kuksa platforms, in which container-based services are deployed in Kubernetes clusters to provide actuation and operation of vehicular systems (see section \ref{sec3C}).\ The service discovery and registry continuously provides a list of available services and registers and deregisters service instances.

The autonomic controller is responsible for the execution of the MAPE-K control loop by using \textit{Monitoring}, \textit{Analysis}, \textit{Planning}, and \textit{Execution} services.\ These are container images that are registered to the device registry mechanism and deployed in orchestration mechanisms, such as Kubernetes to work as the control loop microservices.\ The monitoring services provide solutions for collecting, filtering, and reporting runtime executions and performance data from the managed resources in vehicles, such as domain-specific services, sensors, CANBus, Lin, and network nodes.\ To define variables to be monitored at runtime, it is necessary to collect quality requirements for the healthy functioning of an automotive system (e.g., prioritised emergency messaging of a vehicle accident) and translate them to runtime variables (e.g., network speed $\geq$1MB/s and processing time $<$1s).\ The variables are associated with an automotive scenario (e.g., the emergency braking distance per speed) or the system's behaviour in that scenario (e.g., system reconfiguration time or response time).\ The monitoring results are affected by sensing issues, such as latency, inaccuracy, or reliability~\cite{esfahani2013uncertainty}.

The values of the monitored variables are sent for runtime analysis services to model the system's operating conditions and provide a perception of the current scenario surrounding the managed system of the vehicle.\ Many techniques and solutions (i.e., machine learning or fuzzy logic) can be used here to learn about the performing environment, such as traffic states in different urban areas and provide more efficient and reliable models that facilitate the prediction of future conditions (e.g., analysing data loads and driver's profiles to optimise services' data consumption).\ Using the microservice architecture assures that the analysis services are highly available to model operating conditions at complex runtime situations of vehicles, although the distribution of microservice may require extra monitoring data to decrease the possibility of wrong perceptions of the current environment scenario and the likelihood of triggering faulty adaptation decisions.\ For example, the adaptive cruise control can receive more data from the vehicle's radar to confirm that it provides reliable results (e.g.\ no distortion is present).\ The results of the analysis services are presented as the adaptation strategies, including creating new strategies or keeping existing ones based on the comparison of the optimum and actual working conditions.\ Adaptation strategies are collected and stored in databases called the adaptation strategy registry.

The adaptation planning services are responsible for making decisions about triggering a new adaptation strategy and sending a set of actions needed to reach the optimum state.\ The decisions can include, for example, adding or removing microservice replicas, triggering a new service while stopping a running one (e.g., disabling the cruise control service while alarming the user), replacing a container image with a newly downloaded service, optimising quality attributes in a list of multiple running automotive services, updating the service deployment processes (e.g., deploy new applications only when there is a high-speed network connection or the vehicle is parked), or a combination of these actions.\ The independency among different microservices makes them a secure alternative for runtime planning of new updates and reconfigurations in vehicular systems.\ The solution space could include inputs from, for example, drivers, application providers, or vehicle manufacturers or be created at runtime by using, for example, artificial intelligence models.\ The analysis and planning services may confront challenges concerning latency or model uncertainty \cite{esfahani2013uncertainty}.

The adaptation execution services control the implementation of a strategy and are able to deploy runtime updates.\ Comparing to monolithic architectures, microservices allow simplifying the solution space as we need to test and execute a smaller set of services.\ The strategy could be directly updated in services at runtime, e.g., infotainment systems, or it may require extra privileges from vehicle manufacturers before deployment, e.g., cruise control systems.\ The knowledge from this phase is stored in the knowledge base to be used in more informed decision making over the lifetime of the system.\ Uncertainties related to execution latency or executor reliability are expected at this stage \cite{esfahani2013uncertainty}.

The final part of the Kuksa\textsuperscript{*} framework is the knowledge base that contains multiple repositories, which provide access to the knowledge and data from microservices according to the interfaces prescribed by the architecture.\ The knowledge base maintains data of the managed automotive system (i.e., configuration data, adaptation models, service data consumption, service configurations) and the environment (i.e., traffic modelling data, network data, connected objects profiles), adaptation goals, and the states shared by other four services in the autonomic controller.\ The challenge with using historical data in future scenarios is the effort to capture the delta in knowledge from prior and posterior states, quality requirements, and quality satisfaction levels.
\vspace{-.1cm}

\section{Experimental Evaluation}
\label{sec5}
\vspace{-.1cm}

\subsection{Experimental scenario}
\label{sec5A}
\vspace{-.1cm}
To show how we can implement the Kuksa\textsuperscript{*} framework, we used a laboratory setting and deployed a container-based control loop on the in-vehicle platform of the Eclipse Kuksa to demonstrate self-adaptation microservices in the automotive domain.\ The experiment was implemented on a mobile robotic device called rover.\ The control loop enabled adaptive services for video streaming from a camera on the rover to the cloud back-end.\ The video received in the cloud was encoded and forwarded to a cloud-native streaming service.\ The streaming service was attached to a storage service running on the cloud back-end that was responsible for archiving the received video packages.\ The evaluation has been done by comparing the system performance under evaluation when we had static and adaptive configurations (see section \ref{sec5B1} for more detail on the configuration settings).\ The null hypothesis of this experiment indicates that there is no significant difference between the performance of the static and adaptive systems under evaluation.

We planned that the self-adaptive video streaming scenario uses three microservices in the control loop to monitor, analyse, plan, and execute new configurations based on the changes in the performing conditions and adjust its performance by adapting the video frame rate and scale according to network speed conditions.\ During the experiment, the rover was constantly moving in a circular path in the lab with a speed of 3 km/h.\ The \textit{connection speed test service} was responsible for the monitoring of the network connection speed in the rover.\ The \textit{user's config.\ service} enabled the registration of new user's configurations over-the-air using a Wi-Fi network.\ The \textit{video adaptation service} took the responsibility of the planning and generating new adaptation strategies according to the network connection speed and the configuration setup received from users.\ The list of adaptation strategies was defined as the microservice's input in the adaptation space (see Table \ref{AdaptationSpace}).\ The \textit{video-streaming services} dynamically executed new adaptation strategies from the adaptation strategy registry repository in case of changing configurations or network load conditions.\ We designed multiple data stores (representing the knowledge-base in MAPE-K loop) to be connected to the microservices, in which it allowed us to minimise failures and the overhead load in our system because microservices could access the historical data in case there were service interruptions, service failures, or no new configurations.\ All services were container-based and were running in a Kubernetes cluster that was responsible for creating managing service replicas.\ In the designed scenario, the frame rate and scale of the video were automatically and dynamically reduced if the load of the network increased and passed a specific threshold.\ The threshold was calculated based on the average upload speed in a period of three hours before the experiment.
\vspace{-.45cm}

\subsection{Experimental setting}
\label{sec5B}
\vspace{-.2cm}
\subsubsection{The in-vehicle platform:}
\label{sec5B1}
We used an open-source mobile robotic car, which has a RoverSense layer designed for in-vehicle communication demonstrations and a motor driver layer (Arduino) on top of a Raspberry Pi 3 Model B (RPi3).\ The rover hardware was simulating an electronic control unit (ECU) in vehicles.\ It runs based on the in-vehicle Kuksa platform, which included AGL as the operating system and an API for handling rover communications.\ The in-vehicle Kuksa platform also runs customised software called roverapp\footnote{https://app4mc-rover.github.io/rover-app/} designed for a Linux-based, embedded single board computer (i.e., RPi3).\ The RoverSense layer enables control of the rover and sends telemetry data from rover sensors and camera data, such as video streaming, infrared proximity, and ultrasound.\ In our experiment, the rover sent the video streams using the real-time messaging protocol (RTMP), which was enabled by the FFMPEG project.\ The list of microservices that were running on the Kuksa in-vehicle platform is as Table \ref{Config}.
\vspace{-.65cm}

\begin{table}[H]
 \caption{Microservices configuration}
 \vspace{-.2cm}
 \centering
 \label{Config}
 \begin{tabular}{p{3cm}p{3cm}p{4.5cm}}
 \hline 
\rowcolor{lightgray}\multicolumn{1}{l}{Microservice} & \multicolumn{1}{l}{Technology} & \multicolumn{1}{l}{Configuration}\\
\hline
Service discovery & Hashicorp & consul discovery \\
\hline
Video adaptation & python & -- \\
\hline
Speed test & speedtest-cli & dynamic IP (DHCP)\\
\hline
Video streaming & FFMPEG & maxrate:3M, bufsize:6M, t:30s\\
\hline
\end{tabular}
\vspace{-.8cm}
\end{table}

Also, we had three configuration settings of `low rate' (LR), `high rate' (HR), and `adaptive' in the adaptation space, as shown in Table \ref{AdaptationSpace}.
\vspace{-.65cm}

\begin{table}[H]
\caption{Adaptation space}
\vspace{-.2cm}
\centering
\label{AdaptationSpace}
\begin{tabular}{p{1.75cm}p{1.5cm}cc}
\hline 
\rowcolor{lightgray}\multicolumn{2}{c}{Configuration setting} & \multicolumn{1}{c}{Frame rate} & \multicolumn{1}{c}{video scale}\\
\hline
\multirow{2}{*}{Static} & LR & 30 & 320:240 \\
\cline{2-4} & HR & 60 & 720:480 \\
\hline
\multicolumn{2}{c}{\multirow{2}{*}{Adaptive}} & 30 & 320:240\\
\cdashline{3-4} & & 60 & 720:480 \\ 
\hline
\end{tabular}
\end{table}

\subsubsection{The cloud back-end:}
\label{sec5B2}
\begin{figure}[t]
\centering
\scalebox{0.43}{\includegraphics{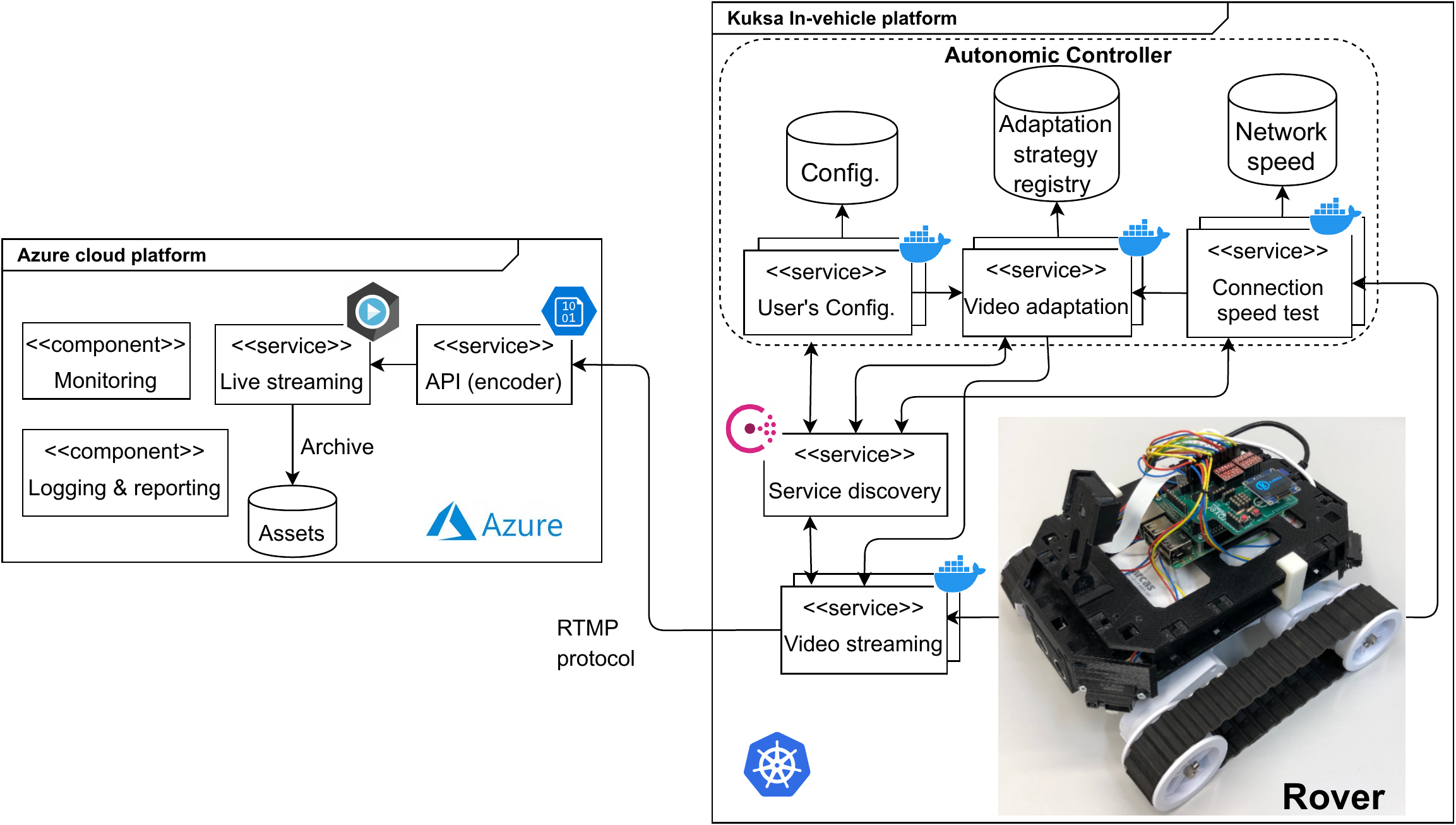}}
\vspace{-.3cm}
\caption{The experimental setting.}
\vspace{-.7cm}
\label{experimentsetting}
\end{figure}

In the designed experiment, the cloud back-end was running on the Microsoft Azure.\ We had a message encoder that was connected to a live streaming service on the back-end.\ The live streaming service was created in ``Media service'' on Microsoft Azure and allowed transforming, analysing, and streaming media on a single platform.\ The videos received in the cloud were archived in a database.\ All communications between the rover and Azure cloud platform were done through an open Wi-Fi connection on the RPi3.\ Figure \ref{experimentsetting} shows the automotive setting in our experiment.

\vspace{-.2cm}
\subsection{Results}
\label{sec5C}
\vspace{-.15cm}
There were three rounds of video streaming from the rover to the cloud back-end.\ The first two rounds were static with fixed parameters for `frame rate' and `video scale', and the third round used the adaptive configuration based on the network load.\ Each experiment round included 100 runs of video streaming with a length of 30 seconds each, meaning 3000 seconds of video streaming in each~round.

\begin{figure}[t]
\centering
\scalebox{0.41}{\includegraphics{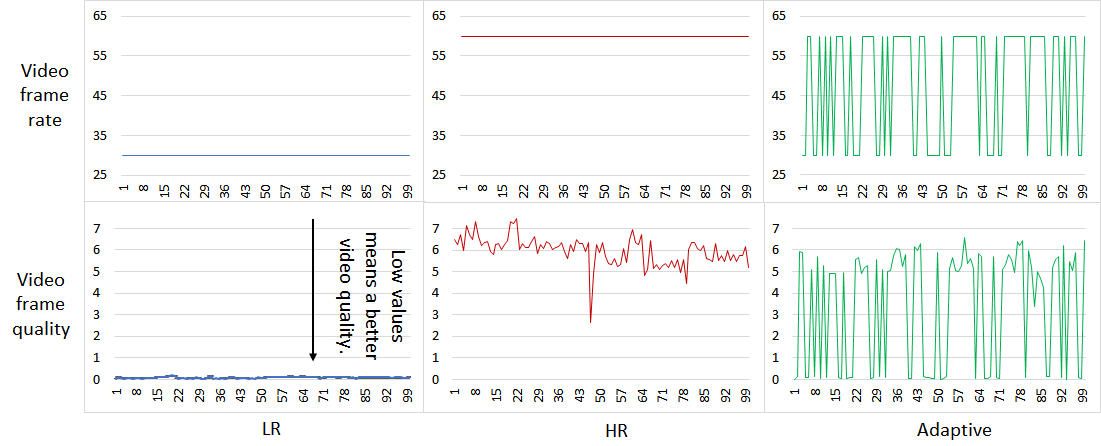}}
\vspace{-.4cm}
\caption{The system behaviour in three configuration settings.}
\vspace{-.7cm}
\label{experimentresult}
\end{figure}

The adaptive system dynamically changed the video streaming configuration according to the latest adaptation strategies.\ The video adaptation service continuously provided new adaptation strategies based on the results from the monitoring and user config.\ services.\ In the adaptive configuration setting, the video adaptation service generated the LR configuration in 31\% of the total runs, and the rest was done with the HR configuration.\ Figure \ref{experimentresult} shows the outcome of our experiment using different configuration settings.\ It shows how the video streaming configuration was changing according to network connection loads.\ It indicates that the adaptive system configuration attempted to reach an optimum for the video frame rate and quality (low values in video frame quality graph means better image quality).

Table \ref{Evaluation} provides the aggregated results of our experiment.\ In this table, $XrYq$ shows quality trade-offs for the video streaming sent to the cloud.\ For example, $5r5q$ declares equal weight for the video frame rate per second (fps) and the video frame quality.\ $9r1q$ shows that the importance of having double fps is nine times more than the frame quality, where $1r9q$ shows the opposite.

On the other side, $p1(sys)$ shows equal weight factors for the time and quality performances ($wt=wq=0.5$), and $p2(sys)$ shows more importance given to the time performance rather than quality performance ($wt=9*wq$), where $p3(sys)$ shows the opposite weight factors and more stress on the quality performance ($wq=9*wt$).
\vspace{-.7cm}

\begin{table}[ht]
\caption{Evaluation results for two architecture designs}
\vspace{-.2cm}
\centering
\label{Evaluation}
\begin{tabular}{p{2cm}p{1cm}p{1cm}p{1cm}p{1cm}p{1cm}p{1cm}p{1cm}p{1cm}p{1cm}}
\hline 
& \multicolumn{6}{c}{Static system}& \multicolumn{3}{c}{\multirow{2}{*}{Adaptive system}}\\
 \cline{2-7} Metric&\multicolumn{3}{c}{LR}&\multicolumn{3}{c}{HR}&&&\\
 \cline{2-10}&5r5q&9r1q&\multicolumn{1}{c|}{1r9q}&5r5q&9r1q&\multicolumn{1}{c|}{1r9q}&5r5q&9r1q&1r9q\\
 \hline
 $tp$&\multicolumn{3}{c|}{1}&\multicolumn{3}{c|}{1}&\multicolumn{3}{c}{0.91}\\
\hline
$qp$&0.74&0.55&\multicolumn{1}{c|}{0.94}&0.6&0.92&\multicolumn{1}{c|}{0.28}&0.68&0.78&0.58\\
\hline
$p1(sys)$&0.87&	0.78&	\multicolumn{1}{c|}{0.97}&	0.8&	0.96&	\multicolumn{1}{c|}{0.64}&	0.8&	0.85&	0.75\\
\hline
$p2(sys)$&0.97&	0.96&	\multicolumn{1}{c|}{0.99}&	0.96&	0.99&	\multicolumn{1}{c|}{0.93}&	0.89&	0.9&	0.88\\
\hline
$p3(sys)$&0.77&	0.6&	\multicolumn{1}{c|}{0.95}&	0.64&	0.93&	\multicolumn{1}{c|}{0.35}&	0.7&	0.79&	0.61\\
\hline
\end{tabular}
\begin{tablenotes}
\item[*] \textbf{Remark:} $tp$: time perf., $qp$: quality perf., $pN(sys)$: three weight settings for time and quality perf., $XrYq$: three weight settings for video frame rate and video frame quality.
\end{tablenotes}
\vspace{-.8cm}
\end{table}

It was necessary to provide multiple sets of trade-off in the evaluation setting because various scenarios in vehicles require customised, adaptive settings according to the system quality requirements; for example, safety-critical scenarios may need higher frame rates or lower configuration times comparing to the vehicular comfort systems.

The results show a better time performance when we had static system, $p2(sys)$, configurations.\ The reason for this is that the configuration time was shorter because the video streaming microservice did not require checking external configuration strategies.\ However, in an adaptive configuration scenario, the video streaming service had to continuously check the new adaptation strategies from video adaptation strategies, in which it reduced the time performance.

The quality performance of the adaptive system stood between the quality performances from the HR and LR configuration settings.\ This means that depending on the importance of the fps or frame quality in our requirements, the adaptive system configuration can be adjusted to deliver the required outcome.\ We could see similar results in the total system performance, $p(sys)$, measurements.\ For example, when we required a higher frame rate with an emphasis on $qp$, the adaptive system showed better results than the static LR configuration setting, but when we needed higher frame quality and higher $qp$, we had better results from the adaptive system compared with the static HR setting.

\begin{tcolorbox}
\textbf{Summary.} To have optimal performance from a self-adaptive automotive system, design trade-offs are necessary at runtime for balancing the expected quality requirements' satisfaction levels of each microservice and the whole system.
\end{tcolorbox}
\vspace{-.1cm}

\section{Discussion}
\label{sec6}
\vspace{-.1cm}

\subsection{Overview of findings and their implications}
\vspace{-.1cm}

\textit{RQ1.\ How can microservices be applied in the design of self-adaptive systems in the automotive domain?}
The vehicular systems should be able to adjust their behaviours according to uncertainties in operating conditions, such as changes in the list of running services, quality requirements satisfaction levels, accessibility to data, adaptability of resources, and working contexts.\ Importing several microservices in the control loop of a self-adaptive system architecture can increase the complexity of the whole system, for example, because of the long reconfiguration time of multiple service replicas or a wrong perception of the vehicle's environment as the reason of conflicting data from different services.\ Although presenting control components in a form of separate microservices to optimise the behaviour of other services increased the modifiability, configurability, and availability of the system when there was a need for reconfigurations.\ System adaptations were isolated to limited services, while other services were not affected.\ This provides higher flexibility in proposing new complex adaptation strategies and models into safety-critical systems of a vehicle.

In vehicular systems, it is necessary to prioritise safety-critical quality requirements over other quality requirements when we make quality trade-off decisions.\ Microservices make it easier to design the safety-critical trade-offs when making adaptation decisions as they minimise the posterior conflicts among services.\ Microservice design allowed us to keep all services lightweight, which reduced the time and effort necessary for reconfiguration.

\textit{RQ2.\ What are the quality trade-offs at runtime when using self-adaptive microservices in the automotive domain?}
We evaluated the runtime performance of a self-adaptive architecture based on microservices in a laboratory setting.\ Our results indicate that the total performance of the system widely depends on the performance of the individual service instances.\ Increasing communication among the multiple service instances increased the total time of reaching the final optimum state after registering a new adaptation strategy, resulting in reduced configuration time performance in the system.

The designed architecture resulted in better fault-tolerance of the system because different service instances could fetch the historical data from the knowledge base in case there was no updated information available.\ This kept the whole system running if the control services could not register adaptation strategies; although multiple data stores may raise challenges, such as data inconsistency.\ Using service management mechanisms, such as Kubernetes improved resiliency and healthy function of all services.

To guarantee performance levels, we designed multiple sets of configuration settings (see Table \ref{experimentsetting}) that may be considered in various operating situations of vehicles.\ The preliminary set of quality attributes and their satisfaction levels for each service and the whole system can be negotiated and designed in vehicular systems; although new configurations may be fetched at runtime from analysis services that use machine learning.
\vspace{-.3cm}

\subsection{Threats to validity}
\label{sec6B}
\vspace{-.15cm}

There are threats to the validity \cite{a2000experimentation} of our results.\ Improving the construct validity required using the right measures in the experiment.\ The measures in this study were composed based on the relevant approaches and metrics for measuring software system performance in peer-reviewed studies.\ Internal validity is addressed through the relationship between the constructs and the proposed explanation.\ We applied an experimental approach in a laboratory setting with defined objectives.\ Also, the study planning and implementation, the experimental setting, and study results were reviewed and discussed among the authors.

External validity is related to the generalisability of the study.\ We decided to use a real-world platform and a mobile robotic device (simulating the vehicular ECUs) to address the adaptive performance of a video streaming service based on time and quality measures.\ Even though the system under evaluation does not represent all scenarios in the automotive domain, such as safety-critical situations, the findings show how self-adaption microservices can be used to adapt to dynamic real-time requirements of vehicular systems.\ In addition, Kuksa\textsuperscript{*} was designed to be flexibly applied in other automotive scenarios.\ However, the findings of this study should not be generalised beyond its original scope.\ We have provided all the details about the experimental setting and all publicly available materials on GitHub\footnote{https://github.com/ahmadbanijamali/Adaptive-video-streaming}.
\vspace{-.3cm}

\subsection{Recommendations for future research}
\label{sec6C}
\vspace{-.15cm}

The measures used in this study were extracted from the prior software engineering studies.\ We propose future studies to investigate the performance of self-adaptive automotive microservices using domain-specific measures from automotive software engineering.\ In addition, future studies could strengthen our results by evaluating our services and experimental setting in a real automotive environment where there exist safety-critical scenarios, such as the service unreliability due to communication network interruptions.\ To improve the understanding of the self-adaptive microservices in automotive systems, we propose more studies on the learning configuration algorithms, domain-specific testing approaches, and quality measurements of self-adaptive microservices.

Adaptive video-streaming services are used as an example to introduce the self-adaption microservice into automotive systems.\ We propose future studies to investigate the application of self-adaptation microservices in other automotive scenarios and review the set decisions about how decentralised each of the MAPE-K services must be made.
\vspace{-.2cm}

\section{Conclusions}
\label{sec7}
\vspace{-.2cm}
Microservices have gained increasing popularity in the automotive domain.\ Considering the dynamic working conditions in vehicles and beyond them, software systems should adapt their behaviours based on changes in the operating and surrounding environments.\ We aimed to investigate self-adaptation microservices in the automotive domain.\ To achieve that aim, we proposed the Kuksa\textsuperscript{*} framework by updating the architecture of a real-world automotive platform and investigated the framework applicability in an automotive laboratory setting.

Our proposed design could adapt the system behaviour according to the changes in the performing conditions.\ Our study showed that system performance requires design trade-offs for balancing the expected quality requirements' satisfaction levels.\ Our findings show the potential of using microservices for self-adaptive automotive systems at runtime.

\end{document}